\documentclass[sigconf]{acmart}

\copyrightyear{2023}
\acmYear{2023}
\setcopyright{licensedusgovmixed}\acmConference[EUROMPI '23]{Proceedings of
EuroMPI2023: the 30th European MPI Users' Group Meeting}{September 11--13,
2023}{Bristol, United Kingdom}
\acmBooktitle{Proceedings of EuroMPI2023: the 30th European MPI Users' Group
Meeting (EUROMPI '23), September 11--13, 2023, Bristol, United Kingdom}
\acmPrice{15.00}
\acmDOI{10.1145/3615318.3615320}
\acmISBN{979-8-4007-0913-5/23/09}

\usepackage{listings}
\usepackage{graphicx}
\usepackage{xcolor}
\usepackage[caption=false]{subfig}

\definecolor{backcolor}{rgb}{0.95,0.95,0.92}
\newcommand\THREAD[1]{\texttt{MPI\_THREAD\_\-#1}}
\newcommand\MPIpThreads{MPI$+$Threads}
\newcommand\MPIxThreads{MPI$\times$Threads}

\newcommand\MPI[1]{\texttt{MPI\_\-#1}}
\newcommand\MPIX[1]{\texttt{MPIX\_Threadcomm\_\-#1}}

\newcommand{\ie}{\textit{i}.\textit{e}.,\ }

\newif\ifdraft
\draftfalse
\ifdraft
  \newcommand{\rajeev}[1]{{\textcolor{red}{ Rajeev: #1 }}}
\else
  \newcommand{\rajeev}[1]{}
\fi

\begin{document}

\title{Frustrated with \MPIpThreads? Try \MPIxThreads!}

\author{Hui Zhou}
\affiliation{\institution{Argonne National Laboratory} \country{Lemont, IL 60439, USA}}
\author{Ken Raffenetti}
\affiliation{\institution{Argonne National Laboratory} \country{Lemont, IL 60439, USA}}
\author{Junchao Zhang}
\affiliation{\institution{Argonne National Laboratory} \country{Lemont, IL 60439, USA}}
\author{Yanfei Guo}
\affiliation{\institution{Argonne National Laboratory}\country{Lemont, IL 60439, USA}}
\author{Rajeev Thakur}
\affiliation{\institution{Argonne National Laboratory}\country{Lemont, IL 60439, USA}}

\begin{abstract}
\MPIpThreads, embodied by the MPI/OpenMP hybrid programming model, is a parallel programming paradigm
where threads are used for on-node shared-memory parallelization and MPI is used for multi-node distributed-memory parallelization.
OpenMP provides an incremental approach to parallelize code, while 
MPI, with its isolated address space and explicit messaging API, affords straightforward paths
to obtain good parallel performance.
However, \MPIpThreads\ is not an ideal solution. Since MPI is unaware of the
thread context, it cannot be used for interthread communication.
This results in duplicated efforts to create separate and sometimes nested solutions for similar parallel tasks.
In addition, because the MPI library is required to obey message-ordering semantics, mixing threads and MPI via \THREAD{MULTIPLE}
can easily result in miserable performance due to accidental serializations.

We propose a new MPI extension, MPIX Thread Communicator (threadcomm), that allows threads to be
assigned distinct MPI ranks within thread parallel regions. The threadcomm extension
combines both MPI processes and OpenMP threads to form a unified parallel environment. We show that
this \MPIxThreads\ (MPI Multiply Threads) paradigm allows OpenMP and MPI to work together in a complementary way to achieve
both cleaner codes and better performance.

\end{abstract}


\maketitle

\pagestyle{plain}
\thispagestyle{plain}

\section{Introduction}
The Message Passing Interface (MPI) has been the backbone of high-performance
computing (HPC) since the mid-1990s. MPI allows an application to run on multiple
compute nodes with distributed memory, thus solving larger problems than what can be solved by using a single node.
MPI can also be used to program multicore machines with shared memory
 by launching multiple MPI processes on each compute node.
Using MPI for both on-node and multi-node parallelization is referred to as MPI-everywhere.
MPI adopts a distributed-process model, where each parallel unit, \ie MPI process,
maintains its own private address space, thus enhancing data locality, preventing accidental synchronization,
and avoiding potential data race conditions.
MPI's explicit messaging API promotes a clean style of
single program multiple data (SPMD) for parallel programming. Its semantic abstraction provides  
straightforward paths to achieve good parallel performance.

However, programming in MPI requires a significant amount of effort to decompose the problem and to parallelize the code.
In contrast, it is incremental, thus easier, to
parallelize code on shared memory by using multithreading techniques, such as OpenMP.
OpenMP allows programmers to start from a single-threaded code and then add
parallel regions to the computation-intensive part via convenient pragma
directives. This ease of use leads 
OpenMP to be the second most used runtime among HPC applications, next to MPI \cite{ecp-usage-20}.
For applications outside supercomputing, multithreading is by far the dominant parallelization technique.

However, the convenience of multithreading does not immediately translate to performance.
The ``naive'' loop-level OpenMP produces frequent entering and exiting parallel regions and many implicit
thread synchronizations, resulting in poor parallel performance~\cite{MPIvOMP-03}.
In order to enhance performance, a significant effort is required to enlarge parallel sections.
The extreme version of larger parallel sections is the SPMD pattern,
at which stage the program become very close to an MPI equivalent code.

Partly due to the significant effort required to transition an OpenMP program
fully into an MPI application and partly due to modern node architectures becoming
more heterogeneous, many applications eventually adopted a hybrid approach to
parallelization: using OpenMP for on-node parallelization and using MPI
for inter-node parallelization~\cite{hetero-18}. For example, the OpenMP 
task offloading facility is a convenient way to program GPUs~\cite{OMP-offload-10}.
This hybrid programming pattern is referred to as \MPIpThreads.

In \MPIpThreads, MPI and threads work independently. To acknowledge this
usage, MPI specifies four thread-compatibility levels:  \THREAD{SINGLE},
\THREAD{FUNNELED}, \THREAD{SERIALIZED}, and \THREAD{MULTIPLE}. The levels
merely specify the thread safety of MPI functions; they do  not pass the
thread execution context to MPI. The most flexible level, \THREAD{MULTIPLE}, requires
nearly all MPI functions to be thread safe. In addition, MPI messages are required
to maintain ordering based on a serial semantic model~\cite{MPI4-threads}.
Without explicit thread execution context,  multithreaded MPI communication
generally has low performance~\cite{mpix-stream}.
Recent studies have demonstrated that scalable \THREAD{MULTIPLE}
performance is possible when an MPI library can correctly map the communication resource
to the application thread context~\cite{Rohit-20}. One way to achieve this mapping is to use dedicated
communicators for each thread (referred to as implicit mapping). In
practice, implicit mapping requires an understanding of implementation detail
and is difficult to rely on. MPICH recently proposed and implemented an
explicit method, MPIX Stream~\cite{mpix-stream}, for applications to pass thread context
explicitly into the MPI library. It has demonstrated reliable and scalable \THREAD{MULTIPLE}
performance.

We recognize that \MPIpThreads\ is  a compromise. Both OpenMP
and MPI are similar runtimes that create a parallel environment and provide facilities
for writing parallel codes. While each has a  different focus---OpenMP focuses on
shared-memory parallelization, and MPI focuses on distributed-memory 
parallelization---both are used for writing similar parallel programming tasks. In \MPIpThreads, similar tasks
require distinct code and separate programming efforts.
For example, an \MPI{Barrier}
synchronizes between processes for the calling thread, whereas an OpenMP barrier
synchronizes only between threads. A global barrier will require a sandwich
call to both barriers.
In addition, both OpenMP and MPI
may introduce compromises. For example, OpenMP code may need manual synchronization before and after
calling MPI because MPI requires the process semantic scope. MPI, on the other hand, may add
thread safety or extra critical sections that can result in unnecessary serialization.

OpenMP and MPI also provide different abstractions and levels of convenience for common parallel tasks.
OpenMP provides dynamic parallelization, allowing programs to enter and exit parallel regions
or even nested parallel regions at will. MPI does offer a dynamic process API, but it is 
difficult to use and is often not optimized. On the other hand, OpenMP
does not offer explicit message-passing semantics and has only minimal collective semantics and
no datatype abstractions. Synchronizing thread-private data and optimizing its performance is often challenging.

Rather than the mutually compromising way of \MPIpThreads, what if we extend MPI and allow it to be
used inside a parallel region directly between threads as well as between processes?
This approach would create a mutually complementary way of using MPI and OpenMP. MPI can utilize OpenMP's
flexibility in creating dynamic parallel regions and expand its parallelization. OpenMP, on the
other hand, can use MPI's explicit messaging, collective APIs, and datatype abstractions to achieve
cleaner code and optimized performance.
We  refer to this new programming pattern as \MPIxThreads\ (MPI multiply Threads).
In an $N$-process MPI program and an $M$-thread OpenMP parallel region, our proposed extension will create
an MPI communicator of size $N \times M$.

In this paper, we introduce the new extension developed in MPICH that supports
\MPIxThreads.
We show our preliminary performance results that match or outperform equivalent codes
in pure MPI or pure OpenMP. More significantly, we demonstrate the new paradigm of parallel
programming where we combine the advantages of both MPI and OpenMP without making compromises.
In Section~2 we describe the proposed APIs as MPIX extensions.
In Section~3 we discuss the implementation details.
In Section~4 we provide three case studies to demonstrate the usefulness of \MPIxThreads.
The first case compares \MPIxThreads\ with the MPI-everywhere model. The second case compares
\MPIxThreads\ with pure OpenMP on a single node. For the third case we look at how \MPIxThreads\ can
be used to call an MPI-centric code, PETSc, in an OpenMP parallel region, and we discuss the lessons learned.
In Section~5 we compare our extensions with previous work.
In Section~6 we discuss the potential impact of the threadcomm extension for MPI users, OpenMP users,
and the field of high-performance computing in general.
In Section~7 we summarize our work.

\section{Proposal: MPIX threadcomm}
We propose a new type of MPI communicator, referred to as MPIX thread 
 communicator, or threadcomm for short, to be used inside a thread parallel region.
A threadcomm is created with the following function.
\begin{lstlisting}[language=C, frame=single, framesep=5pt]
int MPIX_Threadcomm_init(MPI_Comm parent_comm,
    int num_threads, MPI_Comm *threadcomm)
\end{lstlisting}
The communicator creation
semantics are similar to \MPI{Comm\_dup}. At the process level, the new threadcomm
duplicates \texttt{parent\_comm} with associated key values and topology information.
The \texttt{num\_threads} parameter indicates the number of threads for the thread parallel region where this threadcomm will be used.
The newly created threadcomm is an inactive communicator
and  cannot be used yet. This semantic is similar to persistent communication
requests, and the function name is designed to follow a similar naming convention, for example, \MPI{Send\_init}.
The only function that can use an inactive threadcomm is the free function.
\begin{lstlisting}[language=C, frame=single, framesep=5pt]
int MPIX_Threadcomm_free(MPI_Comm *threadcomm)
\end{lstlisting}
Both \MPIX{init} and \MPIX{free} can  be used only
outside thread parallel regions by the main thread. Both functions are collective over 
the parent communicator.

The threadcomm can be activated inside a thread parallel region
with the local number of threads matching the creation parameter. To activate it, one calls the following.
\begin{lstlisting}[language=C, frame=single, framesep=5pt]
int MPIX_Threadcomm_start(MPI_Comm threadcomm) 
\end{lstlisting}
An activated threadcomm must be deactivated before exiting the thread parallel region.
To deactivate it, one calls the following.
\begin{lstlisting}[language=C, frame=single, framesep=5pt]
int MPIX_Threadcomm_finish(MPI_Comm threadcomm) 
\end{lstlisting}
Both \MPIX{start} and \MPIX{finish} are collectives in the threadcomm.
Every thread in the thread parallel region from every MPI process
in the parent communicator needs to call the functions.

An activated threadcomm can be used in MPI communication. Individual threads are assigned
unique ranks and act as if they are MPI processes. If there are $N$ processes in the parent communicator and each contributes $M$ threads,
 the threadcomm will have a size of $N \times M$.
Note that a uniform number of threads per process is not a requirement.
The ranks for the threads are ordered at the process level according to the process rank in their parent communicator.
But within the same process, the rank order between threads is implementation dependent.
In our current implementation, threads are ordered according to their arrival time
in \MPIX{start}. The ranks assigned in a threadcomm are not expected to be persistent if the threadcomm is deactivated and activated again.
An activated threadcomm is restricted inside a single parallel region. If the application spawns a nested parallel region or forks
additional threads, the threads in the nested parallel region or the newly forked threads do not inherit the threadcomm.
With these restrictions, a threadcomm rank uniquely identifies a single thread execution context.

\begin{figure}
    \centering
    \includegraphics[width=\columnwidth]{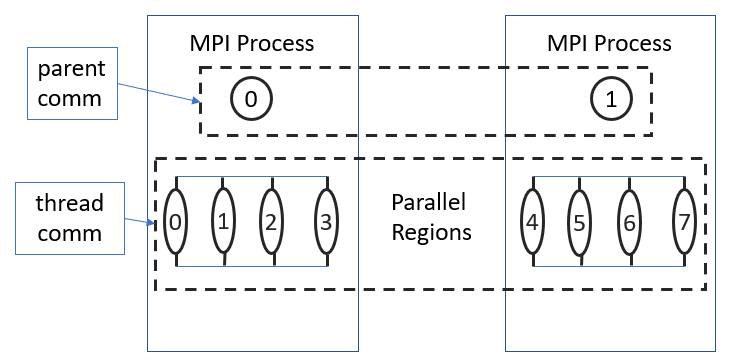}
    \caption{Diagram showing a threadcomm  created in a multithreaded
             parallel region with a size  $N \times M$, where $N$ is the number of
             ranks in the parent communicator and $M$ is the number of threads in each process inside
             the parallel region.}
    \label{fig:threadcomm}
\end{figure}

Listing~\ref{example_threadcomm} and Listing~\ref{example_run} show an example.
Figure~\ref{fig:threadcomm} illustrates the relationship between a threadcomm and its parent
communicator.

In the example code, we do not request  \THREAD{MULTIPLE} even though we are calling MPI from the parallel region.
This is because, inside the parallel regions, all calls to MPI are using the threadcomm context.
The MPI thread level is an assertion passed from the application to the MPI library because the library normally
cannot tell the thread context and thus relies on the asserted thread level to execute correctly and optimally.
However, with threadcomm, the thread contexts are uniquely identified by the threadcomm rank, thus the library
no longer needs the user-asserted thread level for correctness and optimization.
The thread level is still relevant for MPI functions that do not use a threadcomm context.
For example, if a datatype creation function is used inside the parallel region where a threadcomm is active,
then \THREAD{MULTIPLE} will be
required, because in such cases the implementation has no way of detecting the thread context
and hence a blanket thread-safety measure must be applied. We recommend moving such codes outside
the parallel region and avoid \THREAD{MULTIPLE} for optimal performance.

\begin{lstlisting}[language=C, showstringspaces=false, label=example_threadcomm, frame=tb, caption=Example using MPIX thread communicator with OpenMP]
#include <mpi.h>
#include <stdio.h>
#include <assert.h>

#define NT 4

int main(void) {
    MPI_Comm threadcomm;

    MPI_Init(NULL, NULL);
    MPI_Threadcomm_init(MPI_COMM_WORLD, NT,
                        &threadcomm);

    #pragma omp parallel num_threads(NT)
    {
        assert(omp_get_num_threads() == NT);
        int rank, size;
        MPI_Threadcomm_start(threadcomm);
        MPI_Comm_size(threadcomm, &size);
        MPI_Comm_rank(threadcomm, &rank);
        printf("  Rank %d / %d\\n", rank, size);

        /* MPI operations over threadcomm */

        MPI_Threadcomm_finish(threadcomm);
    }

    MPI_Threadcomm_free(&threadcomm);
    MPI_Finalize();
    return 0;
}
\end{lstlisting}

\begin{lstlisting}[language=sh, label=example_run, frame=tb, caption=Running the thread communicator example code]
$ mpicc -fopenmp -o t t.c
$ mpirun -n 2 ./t
    Rank 4 / 8
    Rank 7 / 8
    Rank 5 / 8
    Rank 6 / 8
    Rank 0 / 8
    Rank 1 / 8
    Rank 2 / 8
    Rank 3 / 8
\end{lstlisting}

The semantics of using threadcomm are generally clear by pretending that threads are MPI processes within the parallel region.
However, some MPI functions have semantics that go beyond a specific communicator.
For example, an MPI group derived from a threadcomm may create confusion outside the parallel region
or when interacting with MPI groups derived from process-level communicators.
Thus, we adopt the principle that all threadcomm-derived objects will have their lifetime within the activation
window of the given threadcomm. For example, one can set an attribute to an active threadcomm, but it will get deleted at \MPIX{finish}.
A similar lifetime applies to an MPI group that is derived from a threadcomm. In this case the user has to call \MPI{Group\_free} before calling \MPIX{finish}.
Unlike MPI-everywhere, in \MPIxThreads, code with process-level semantics can be
arranged outside the thread parallel regions; thus most complications from mixing threadcomm context and process-level context can be avoided.

\section{Implementation}
We implemented our proposal in MPICH as a set of MPIX extensions.
This feature is currently available in MPICH's main development branch and
will be released in MPICH version 4.2. Currently implemented threadcomm features
include blocking point-to-point communication, nonblocking point-to-point communication,
and blocking collectives. Additional features such as one-sided communications are
planned for future releases.
In this section we discuss some of the implementation details.
For ease of reading, we will use the term "rank" broadly to denote both a process within a regular MPI communicator and a thread within a threadcomm communicator.

\subsection{Shared memory and thread-local storage}
An MPI communicator internally holds a few types of data.
The first type of data is common for all ranks in the communicator, such as
the communicator size, context id, and rank table. The rank table allows one rank
to look up the network address of another rank. The common data is initialized during
communicator creation and remains read-only afterward.
In the distributed process model, common data is duplicated in every process of
the communicator because of private memory space between processes.
With threadcomm, only one copy of the data per process is needed, thanks to
the automatic shared-memory space between threads.
The second type of data is the shared-memory communication device.
We can think of this data as an array of mailboxes, one for each rank or each rank-pair.
The sender fills the receiver's mailbox with messages, and the receiver consumes the messages.
We will describe shared-memory messaging in the next subsection.
The third type of data is rank-specific. This includes the local rank, posted receive
queue, and unexpected message queue. For threadcomm, rank-specific data needs to reside in
thread local storage (TLS).

Both the common data and shared-memory communication device can be initialized
outside the parallel region in \MPIX{init}. Rank-specific data can only be created in
\MPIX{start} inside the parallel region when thread local storage become available.

\MPIX{init} is a heavy call. Not only does it need to duplicate the parent communicator, but it also
needs to run allreduce on the number of threads in order to populate the rank table.
On the other hand, \MPIX{start} is a relatively lightweight call,
since all it needs to do is to initialize the TLS fields and it is local at the process level. 
Thread ranks are established by atomically
incrementing a shared sequence number and taking its value as its local thread id.
The local thread id can be translated into the rank in the threadcomm based on the rank table.

Existing code in MPICH needs to be patched to check whether an input communicator is a
threadcomm, and it requires a separate implementation to handle a threadcomm.
For point-to-point functions, we implemented new algorithms to handle
interthread messaging. Individual thread ranks can also send messages to another thread rank between different processes.
These interprocess messages are dubbed
into regular point-to-point messages by adding extra bits to the internal tag
and adjusting the source and destination ranks appropriately.
Thread safety needs to be added to avoid race conditions to the network resources.
Alternatively, each thread can be assigned a distinct virtual communication interface
(VCI)\cite{mpix-stream} to preserve thread-concurrency.

Most collective algorithms consist of internal point-to-point communications,
and they can work without modification as long as we patch the macro to
use the correct ranks for the threadcomm.
However, one can implement threadcomm-aware algorithms that take
advantage of the shared-memory space to achieve better performance. We will
look at an example in a case study below.

Unpatched functions may incorrectly treat the threadcomm as a regular communicator. In addition to
incorrect semantics, it may cause race conditions due to concurrent access to shared data.
Until these functions are patched, we place assertions to prevent accidental usage.

\subsection{Shared-memory messaging}
The algorithms to realize interthread messaging are similar to shared-memory-based intranode messaging. We adopted a lockless multiple producers single consumer (MPSC)
queue algorithm that is currently used by MPICH intranode messaging.
For  details about the lockless MPSC queue, see ~\cite{nemesis-06}.
In shared memory, we maintain $N$ lockless queues, one for each
thread rank. A send operation will allocate a cell from the shared memory,
load the cell with the message envelope and the message data, then enqueue to the receiver queue.
Multiple sender ranks may concurrently enqueue to the destination queue,
but only the receiver rank will dequeue from its own queue.

With interprocess shared-memory messaging, allocating shared memory requires collective
communication. Thus, it is often implemented with shared-memory pools that are preallocated
during \MPI{Init}. In order to avoid congestion, the shared-memory pools are also divided among ranks.
The sender will obtain a cell from the receiver cell pool, and the receiver will return the cell to
its own pool once the message is consumed.
The shared-memory pool also restricts cell size. Larger messages that do not fit into a single cell
have to use a pipeline algorithm that uses multiple cells per message.
In addition, the sender needs to accommodate the possibility of the shared-memory pool running out of cells
and have a mechanism to postpone the send.

With interthread messaging, the entire memory space is automatically shared. Thus the cell pool
allocation becomes optional. The sender can directly allocate memory to be used for cells with flexible
cell sizes and enqueue to the receiver's MPSC queue. The receiver can directly free the
cell once the message is consumed. Thus, the code can be much simpler than the equivalent code for
interprocess messaging. However, the performance of messaging may depend on the performance of multithreading
memory allocation, and we found the Linux libc malloc unsatisfactory.
For more predictable performance, we opted to use a shared-memory pool for cell allocation.
This shared-memory pool is initialized during \MPIX{Init}.

When the message size is small and fits inside a single cell,
we simply copy the message into the cell and complete the
sender side. This is referred to as eager mode. The eager mode provides better latency since the sender side does not need
to wait for the receiver side to complete the send.
When the message size is large, we can first send a header to the receiver and wait for an acknowledgment from the receiver
before sending the data. This is referred to as rendezvous (rndv) mode.
Alternatively, when the receiver can directly access the sender buffer, we may send the header
and have the receiver directly copy the data from the sender buffer into the receiver buffer.
This is referred to as a 1-copy algorithm, while both the eager and rndv are referred to as 2-copy algorithms.
In order to implement an interprocess 1-copy algorithm, the sender buffer needs to be
mapped to the receiver side's virtual address space. This mapping has
significant overhead, making the algorithm suitable only for very large messages,
although a caching mechanism can be used to amortize the cost.
In contrast, threads share address space
automatically, and hence we can skip the address mapping overhead altogether.
Thus, large interthread  messages are implemented entirely by using the 1-copy algorithm;  rndv is not used.

\section{Case studies}
In this section we look at some experiments and explore some use cases for threadcomm.

\subsection{Case study: point-to-point latency and bandwidth} \label{case1}
\begin{figure}
    \centering
    \includegraphics[height=2.25in]{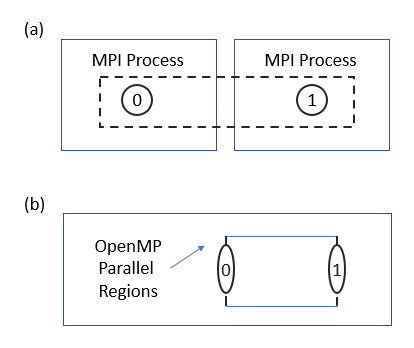}
    \caption{Diagram showing the setup for the point-to-point case study.
             (a) Launch multiple MPI processes on a single node.
             (b) Launch a single process, use OpenMP to create a parallel region, and use threadcomm for MPI point-to-point messaging.}
    \label{fig:case1}
\end{figure}

In this case study we compare the performance of using MPI via the thread
communicator versus MPI-everywhere, where we launch multiple MPI processes on a single node.
See Figure~\ref{fig:case1}.

\begin{figure}[htp]
    \centering
    \includegraphics[clip,width=\columnwidth]{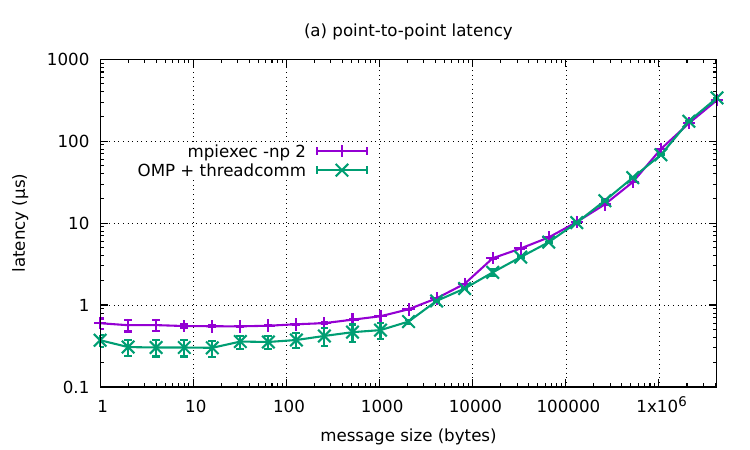}
    \includegraphics[clip,width=\columnwidth]{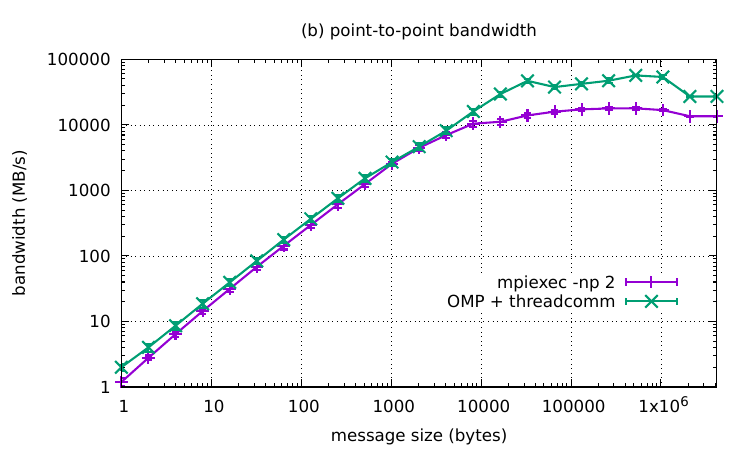}
    \caption{Point-to-point message latency and bandwidth comparison between MPI-everywhere and OpenMP+threadcomm on an Intel Xeon Gold 5317. Processes or threads are bound to cores on the same socket.}
    \label{fig:omp_latency}
\end{figure}

We measured the point-to-point communication latency and bandwidth comparing MPI-everywhere
and using threadcomm within an OpenMP parallel region.
The results are shown in Figure~\ref{fig:omp_latency}.
The interthread messaging via threadcomm shows better latency for small messages and
higher bandwidth for large messages. For this evaluation, the interthread messaging switches from the eager algorithm
to the 1-copy algorithm at the threshold of 4096 bytes. The interprocess shared-memory messaging switches from
the eager algorithm to the rndv algorithm at the threshold of 16 kilobytes. The current MPICH release supports only 
1-copy algorithm via XPMEM, which is not enabled on the evaluation machine.

The shorter latency for small messages can be attributed to a shortcut that allows us to skip
the allocation and deallocation of sender request objects. A request object is needed for tracking when the message
does not fit inside a single cell or requires synchronization from the receiver side such as in
the 1-copy algorithm. Skipping the request object for small messages reduces the instruction count and
improves the latency. This optimization in principle can be applied to interprocess shared-memory messaging as well,
but it is more difficult because of the current MPICH design. For larger messages, the interthread messaging uses the 1-copy
algorithm, and interprocess messaging uses the 2-copy rndv algorithm. Both show similar latency.
For small messages, interthread messaging shows slightly higher bandwidth, which can be attributed to better latency.
For large messages, the much higher bandwidth is due to the 1-copy algorithm. The bandwidth drops above the 1 MB message size
because of heavy last-level cache and TLB misses.

The better performance of threadcomm interthread messaging demonstrates that
launching a single process on each compute node and using OpenMP to launch parallel regions
is an effective alternative to MPI-everywhere. The code inside the parallel region can still
benefit from variable privatization via function abstractions. Thus we expect minimal code changes
with matching or even improved performances.

For OpenMP programs, privatization is one of the key steps in optimizing parallel performance.
However, OpenMP lacks facility for synchronizing between private data.
We expect the threadcomm messaging facility will supplement OpenMP and thus reduce the programming effort during
optimization.

\subsection{Case study: collectives}
In this case study we compare OpenMP+threadcomm with vanilla OpenMP. In
particular, we look at one of the main features offered by MPI, the collectives.
MPI collectives are a high-level abstraction for communication operations in which  a group
of processes need to exchange messages. Rather than explicitly sending
and receiving such messages, MPI provides 17 collectives, not counting nonblocking and
neighborhood collectives, for programmers to write clean SPMD-style parallel code.
The abstraction also separates the optimization effort from application programmers to
runtime developers, improving the performance of parallel code in general.

OpenMP provides some basic collectives via pragma directives. For the thread
barrier, there is \texttt{\#pragma omp barrier}. For simple reduction,
there is the reduction clause. For general collectives over more
complex data, for example, arrays, more tedious manual effort is often required.
The threadcomm extension offers an alternative of using MPI instead.

\begin{figure}
    \centering
    \includegraphics[width=\columnwidth]{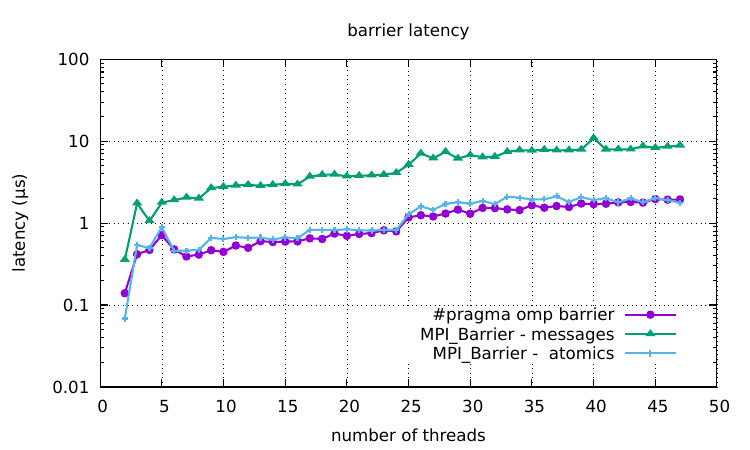}
    \caption{Latency comparison between the OpenMP barrier and MPI\_Barrier via
             the thread communicator on an Intel Xeon Gold 5317.
             Threads are bound to cores.}
    \label{fig:omp_barrier}
\end{figure}

Listing~\ref{lst:barrier} shows an example of replacing \texttt{\#pragma omp barrier}
with \MPI{Barrier}. Figure~\ref{fig:omp_barrier}  presents the measurement results.

\begin{lstlisting}[language=C, label=lst:barrier, frame=tb, caption=Example code replacing OpenMP barrier with \MPI{Barrier}]
#pragma omp parallel
{
    MPI_Threadcomm_start(comm);
  #ifdef USE_MPI
     MPI_Barrier(comm)
  #else
     #pragma omp barrier
  #endif
    MPI_Threadcomm_finish(comm);
}
\end{lstlisting}

The \MPI{Barrier} in MPICH is implemented by using the dissemination algorithm \cite{dissemination-88}.
The existing code implements the algorithm using point-to-point messages. Each MPI
process sends and receives $lg N$ number of zero-sized messages, where $N$ denotes the
number of processes.
The same code can be applied to the threadcomm barrier as long as the communicator rank and size
are retrieved correctly via updated macros. Most existing collective algorithms can work with
threadcomm this way as long as point-to-point communication works.

As shown in Figure~\ref{fig:omp_barrier}, the performance of \MPI{Barrier} using point-to-point messages
is worse than that of the OpenMP native barrier. This is because OpenMP barrier is typically implemented by using
shared atomic variables. In comparison, MPI messaging involves assembling message envelopes, enqueueing
and dequeuing to message queues, and message matching. The disadvantage of messaging versus using atomics is exaggerated in the barrier case where the messaging has no payload. We'll see the advantage disappear when
there is payload in the next reduction example.

Nevertheless, we want to show that a good abstraction is worthwhile to have even if the implementation is
temporarily lagging. In principle, OpenMP barrier and \MPI{Barrier} have the same abstraction within the shared-memory domain, and
the performance difference is only a matter of implementation. Realizing that MPICH's code is not taking
advantage of the shared address space between threads, we reimplemented the same dissemination algorithm
using shared atomics. As shown in Figure~\ref{fig:omp_barrier}, the performance then matches that of the OpenMP barrier.
Readers should note that MPI collectives work across unified parallel environment both within and across
processes. In \MPIpThreads, a global barrier only can be achieved with a thread barrier, then
an MPI barrier from a single thread, followed with another thread barrier.
A single \MPI{Barrier} using \MPIxThreads\ is much cleaner.

\begin{figure}
    \centering
    \includegraphics[width=\columnwidth]{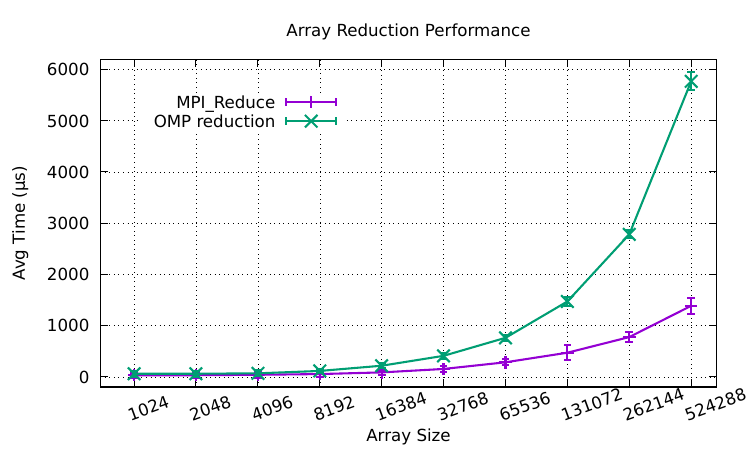}
    \caption{Latency comparison between OpenMP reduction and MPI\_Reduce via
             threadcomm on an Intel Xeon Gold 5317,
             using 16 threads bound to cores.}
    \label{fig:omp_reduce}
\end{figure}

OpenMP also has a reduction clause to aggregate private copies of data. This roughly corresponds
to the semantics of \MPI{Reduce}, although \MPI{Reduce} can be used inside the parallel region on demand
without introducing extra entering and exiting parallel regions while OpenMP's reduction is per parallel region.
Listing~\ref{lst:reduce} shows an example where we reduce an array of integers.
Figure~\ref{fig:omp_reduce} shows the latency measurements. The measurements include the overhead of entering
and exiting parallel regions and the initialization of private array data.
The private array is explicit in the MPI case but implicit in the OpenMP case.
\MPI{Reduce} uses MPICH's stock
``binomial'' algorithm without modification.

The results show \MPI{Reduce} performing much better than the OpenMP reduction. The performance is specific
to the MPI and OpenMP implementations, in this case, MPICH and gcc. For a given example, there is nothing
to prevent OpenMP to adopt the exact same algorithm as MPI or vice versa. However, the reduction clause in
OpenMP is based on language semantics and is more general, while MPI uses more specific datatype and reduction-op abstractions, which allows an implementation to better optimize.

\begin{lstlisting}[language=C, label=lst:reduce, frame=tb, caption=Example code comparing OpenMP reduction with \MPI{Reduce}]
int sum[N];
#ifdef USE_MPI
  #pragma omp parallel
  {
    MPI_Threadcomm_start(comm);
    int my[N];
    int tid = omp_get_thread_num();
    for (int i = 0; i < N; i++) my[i] = tid;
    MPI_Reduce(my, sum, N, MPI_INT, MPI_SUM, 0, comm);
    MPI_Threadcomm_finish(comm);
  }
#else
  #pragma omp parallel reduction(+:sum[:N])
  {
    int tid = omp_get_thread_num();
    for (int i = 0; i < N; i++) sum[i] = tid; 
  }
#endif
\end{lstlisting}

\subsection{Case study: using PETSc}
For the third case study we explore using PETSc \cite{petsc} with OpenMP.
PETSc, the Portable, Extensible Toolkit for Scientific Computation, is designed for the scalable solution of scientific
applications modeled by partial differential equations. Among a lot of functionalities, it provides a rich set of linear solvers,
nonlinear solvers, and optimizers.
PETSc is widely used in academia and industry and has been
evolving into an ecosystem including applications, libraries, and frameworks built on PETSc.
PETSc was originally written in MPI. There were efforts to add \MPIpThreads\ hybrid parallelism
into PETSc \cite{lange2013achieving}; but because of the extra code complexity and often inferior performance compared
with the MPI-everywhere mode, PETSc later removed the hybrid support and has been using only MPI since then.
For users wanting to use PETSc with threads, if their code supports MPI, PETSc would suggest adopting
the MPI-everywhere mode. Otherwise users can only use PETSc in a strict case, where each thread creates sequential PETSc
objects on \texttt{MPI\_COMM\_SELF} and solves small and isolated problems. 
The usability of this case is very narrow.

Recently,  another effort, called PCMPI, was undertaken to support broader OpenMP PETSc users.
Users have to run their OpenMP code in the MPI style, namely, ``\texttt{mpiexec -n np ./myapp}''.
Users must call \texttt{PetscInitialize()} at the
 beginning of their code. Only rank~0 exits the function and continues with the user's OpenMP code; other ranks are in a loop waiting for commands and data from rank~0.
With the option \texttt{``-pc\_type mpi''}, rank 0 will create an MPI parallel Krylov solver (KSP) from a sequential preconditioner (PC)
that lives on rank 0. At the beginning, rank 0 scatters PETSc matrices and vectors to other ranks.
Then all ranks do the solve in parallel. At the end, rank 0 gathers results. The gather/scatter steps are also done with MPI.
One can see that this approach is not flexible since the number of MPI ranks is fixed, and PETSc needs special handling to free
CPU cores occupied by idle MPI ranks to OpenMP threads spawned by users on rank 0.

With threadcomm, PETSc could support OpenMP users more naturally. 
The style is similar to PCMPI described above, but 
users can run their code in a natural OpenMP pattern. Whenever the code
need to call PETSc solvers, they can create a threadcomm then create PETSc objects and solvers on the communicator within
an OpenMP parallel region.  PETSc does parallel computation on the communicator as if it was in a pure MPI environment.
Users still need to manage scattering data from outside the parallel region into the parallel region
where PETSc uses distributed data structures. A future threadcomm-aware PETSc should provide new APIs to facilitate this task.
Listing~\ref{lst:petsc} shows a skeleton code using PETSc with OpenMP.

\begin{lstlisting}[language=C, label=lst:petsc, frame=tb, caption=Example code using PETSc with OpenMP]
    int nthreads = 4;
    MPI_Comm comm;

    MPI_Init(NULL, NULL);
    PetscInitialize(&argc, &argv, NULL, NULL);
    MPIX_Threadcomm_init(MPI_COMM_WORLD, nthreads,
                         &comm);
    #pragma omp parallel num_threads(nthreads)
    {
        assert(omp_get_num_threads() == nthreads);
        Mat A;
        MPIX_Threadcomm_start(comm);
        MatCreate(comm, &A);
        /* 
         * Build matrix A with data from outside
         * the parallel region and perform 
         * parallel computations.
         */
        MatDestroy(&A);
        MPIX_Threadcomm_finish(comm);
    }
    MPIX_Threadcomm_free(&comm);
    PetscFinalize();
    MPI_Finalize();
\end{lstlisting}

We now discuss our preliminary experiments in carrying out this strategy.
Our original attempts to call PETSc entirely within a single OpenMP parallel region
were unsuccessful. The reason is that PETSc assumes a distributed process environment
and thus freely uses global variables, which are private per rank in a conventional MPI
environment. This is not true inside the OpenMP parallel region, and calling PETSc causes
race conditions.
One solution is to turn all PETSc global variables into thread-local storage. 
Doing so, however, will limit PETSc usage within a single OpenMP parallel region, and it is a rather
stringent limit because most OpenMP programs do not have only a single parallel region.
Through investigation, we discovered that most global states in PETSc are initialized in
\texttt{PetscInitialize} and largely read-only during typical usage. Thus,
we can call \texttt{PetscInitialize} and \texttt{PetscFinalize} outside the
parallel regions. In fact, this approach makes sense since  maintaining a duplicated
copy of shared global data is unnecessary and wasteful.
The threadcomm is attached to the matrix created via \texttt{MatCreate}; thus, we need to make sure the matrix
is destroyed before we deactivate the threadcomm and exit the parallel region. PETSc needs to carefully reference
count the Matrix objects and make sure to release all comm-related resources when all the referenced objects are destroyed
and go out of scope.

PETSc also uses global states for logging and debugging. Since
these are all optional features, we simply disabled them for this study. A threadcomm-aware
PETSc should apply thread-safety for logging or use thread-local storage instead.
Inside the parallel region, PETSc parallel matrices and vectors as private variables
and they are constructed from scratch binding to a given communicator. As long as all these PETSc
objects are private to the thread and have their lifetimes within the parallel region, 
they can be used without any changes from an MPI-only code.

One MPI feature that PETSc heavily uses is communicator attributes. We added 
support of the threadcomm attribute via thread-local storage, thus allowing different threads
to concurrently set and retrieve their own attributes.
PETSc also duplicates the user's input communicator internally to avoid conflicts with the user's own usage.
This is also supported in our implementation. The duplicated threadcomm is born as an active
threadcomm and it has to be used and freed within the same parallel region.

\begin{figure}
    \centering
    \includegraphics[width=\columnwidth]{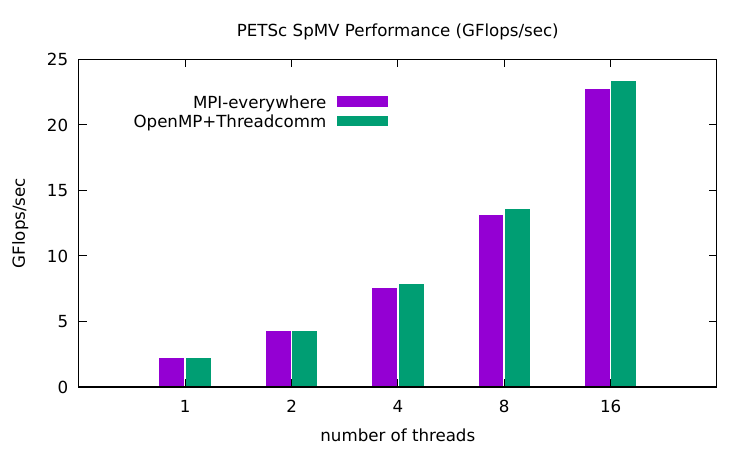}
    \caption{PETSc MatMult (SpMV) performance comparison between MPI-everywhere and OpenMP + threadcomm. The sparse matrix was obtained from a 27-point stencil code on a cube of size 128.}
    \label{fig:omp_petsc}
\end{figure}

Figure~\ref{fig:omp_petsc} shows the PETSc matrix-vector multiplication (MatMult) performance with MPI-everywhere
and OpenMP $+$ threadcomm. 
The experiments were done on a machine with two Intel Xeon 5317 CPUs at 3.00GHz with total 24 cores.
Both show good scalability. The threadcomm version shows slightly better
performance, which is in agreement with the point-to-point performance results as in case study \ref{case1}.

\section{Related work}
Here we discuss some related work in this area.

\subsection{Special MPI implementations based on threads}
A number of previous efforts implement MPI using threads or thread-like processes
for MPI processes.

Fine-grain MPI (FG-MPI) \cite{fg-mpi} implements MPI processes on proclets, a version of user-level threads.
A proclet yields to the scheduler whenever the proclet blocks on an MPI communication call.
MPC-MPI \cite{mpc-mpi} implements MPI on Multi-Processor Computing (MPC) runtime, which supports MPC tasks, another version of user-level threads.
MPC runtime supports both MPI and OpenMP but only  as \MPIpThreads; that is, OpenMP threads
share the fixed MPI process.
User-level thread runtimes offer low context-switch cost and thus can support running many
MPI processes even in an oversubscribed situation.

Both FG-MPI and MPC-MPI are MPI implementations that run on special runtime systems.
Both implementations target classical MPI programs, namely, the SPMD paradigm.
On the other hand, threadcomm is proposed as an extension to the MPI standard.
We are targeting both MPI and OpenMP programmers and proposing a new \MPIxThreads\ programming paradigm.

MPICH has also been extended to process-in-process (PiP)\cite{pip-18} runtime, where MPI
processes are implemented as PiP processes. PiP processes share address space, but
each process still maintains its own private variables.
PiP-MPICH is yet another thread-like MPI implementation targeting MPI applications.
PiP-MPICH is able to take advantage of the automatically shared address space and implement
special algorithms, such as work stealing, for faster communication~\cite{cabmpi-20}.

While threadcomm shares some implementation details with these special MPI implementations,
such as interthread messaging, 1-copy algorithm, and the use of thread local storage,
our goal is not just to develop a special implementation with enhanced performance for existing
MPI applications. Rather, our goal is to provide new facilities for applications to explore
a new \MPIxThreads\ style of parallel programming, which we believe is both easier to program and better
in performance.

\subsection{Comparison with the MPI endpoints proposal}
The MPI endpoints proposal~\cite{endpoints-14}
proposed a single API extension to address the dynamic
threading environment.

\begin{lstlisting}[language=C, frame=single, framesep=2pt]
int MPI_Comm_create_endpoints(MPI_Comm comm, int num_ep, MPI_Info info, MPI_Comm comm_out[])
\end{lstlisting}

This API is similar to \MPIX{init}. A notable difference is that  \MPIX{init} outputs a
single threadcomm handle, whereas \MPI{Comm\_create\_endpoints} outputs
an array of communicator handles, each handle corresponding to a single endpoint rank.
The reason for outputting an array of communicator handles is that the function is
invoked outside the thread contexts and there is no implicit thread local storage to hold
rank-private information. By returning an array of communicator handles, each handle
is associated with a single rank and can hold rank-specific data.
Upon entering the thread parallel region, it is expected that each thread be assigned
one of the handles, thus assuming a unique rank within the endpoints communicator.
The obvious drawback of the endpoints proposal is that it burdens the users with managing
rank-private communicator handles outside the parallel region where the thread context
does not exist yet. While technically this is only a chore, it is conceptually confusing
and is susceptible to abuse.
In fact, while the endpoints communicator is expected to be used inside a thread parallel region,
the proposal places no restrictions on different usages. For example, it allows a single thread
to assume multiple thread ranks or even to switch thread ranks. This makes it
 difficult to implement the endpoints proposal. Many complications can arise from
unexpected use cases.
Fundamentally, the endpoints proposal did not address the issue that MPI is not aware
of the thread context. An MPI implementation may need to ensure progress on
all endpoints of a process regardless of which endpoint rank invokes the progress.
This can easily result with a severe performance penalty when multiple threads concurrently
invoke progress.

The threadcomm extension, on the other hand, restricts the usage to be only inside a parallel region, where
thread-local storage can be used by implementation to implicitly track the thread context.
In a way, MPI becomes thread-aware.
The implementation does not just treat individual thread rank as an 
individual process; it can treat the thread rank as what it is, a thread.
It can track rank-private data, afford cleaner syntax for user code, and use shared memory to coordinate and optimize algorithms, thereby achieving better performance.

\section{Discussion and Perspective}

Here we  discuss the potential impact of the threadcomm
extension.
For experienced MPI users, threadcomm offers an alternative to MPI-everywhere.
We can view OpenMP as an in-process launcher that can dynamically create and shrink
MPI processes. This is traditionally addressed by the MPI dynamic process
API, namely, \MPI{Comm\_spawn} and family. However, the implementation of MPI dynamic
processes is challenging and  is lagging currently in most MPI implementations.
In particular, spawned processes on the same node often cannot use shared memory
for communication because of the missing coordinated setup stage.
Comparing the dynamic processes with OpenMP $+$ threadcomm, we find it  easy to spawn
dynamic parallel regions that effectively expand the MPI parallel environment. Moreover, the narrow
context of threadcomm makes it easy to take advantage of the thread
context and provide a high-performance implementation.

For experienced OpenMP users, a typical strategy in performance optimization is
via privatization, where dependency between threads on shared variables is removed by
adding a private copy of the data, so the thread can read and write to the data
independently and thus in parallel. The privatization immediately brings a challenge: 
How do you synchronize private data? For this, the facility OpenMP provides is very limited.
On the other hand, MPI is designed for this and offers a plethora of APIs, from point-to-point
send and receive to collectives, from blocking to nonblocking, and MPI datatypes for
describing the private data layout.
Thus, the threadcomm extension complements OpenMP by providing OpenMP with much-needed messaging and synchronization facilities.

For general HPC application developers, there is often a natural progression
from a single-threaded prototype, where the focus is the science and correctness,
to a strenuous parallelization effort, where the focus is scaling and performance.
The latter is  challenging for domain scientists. In a way, OpenMP is much easier
than MPI because it allows the developer to retain some of the single-threaded code and
focus on parallelizing computation-critical parts one at a time. At some stage, when
OpenMP on a single workstation can no longer accommodate the scale, the application
often requires an overhaul to migrate to MPI.
The threadcomm extension introduces MPI to application developers at a much earlier stage,
even when the application is still parallelizing within a single node. Since the principle
of high-performance parallel computing is the same within a node or within a cluster,
we believe introducing MPI early  will allow a smoother development cycle.

We observe that HPC researchers often inadvertently build silos. This is  natural 
as the researchers dig more deeply into their field and the field become narrower as a result.
Despite the similar goals, challenges, and principles, MPI research and OpenMP
research nevertheless form separate communities. And solutions rarely commute. For example, MPI cannot be
used between threads, and OpenMP cannot be used to spawn MPI processes. OpenMP needs to figure
out its synchronization and off-node solutions despite the fact that MPI has provided them all along.
Vice versa, MPI needs to figure out how to effectively spawn threads and take advantage of the
shared memory despite the fact that OpenMP already does that. 
We believe that the threadcomm extension can be the
missing link to bridge the gap between MPI and OpenMP.

\section{Summary}
In summary, we propose and have implemented in MPICH the threadcomm extension to allow multithreaded
applications to directly use MPI inside parallel regions.
Unlike in \MPIpThreads, where MPI and threads form disjoint parallelizations,
threadcomm multiplies the number of processes and the number of threads per process
and forms a simpler, larger parallelization, thus a new \MPIxThreads\ paradigm.
We show that threadcomm enables dynamic expansion of MPI processes and performs better
than MPI-everywhere by taking advantage of shared memory.
It also brings the MPI facility to OpenMP and can ease the development cycles for
HPC applications.

\begin{acks}
We  thank Jed Brown for not giving up on the endpoints proposal and the
discussions that led to this work.
We gratefully acknowledge the computing resources provided and operated by the
Joint Laboratory for System Evaluation (JLSE) at Argonne National Laboratory.
This research was supported by the Exascale Computing Project (17-SC-20-SC), a
 collaborative effort of the U.S.\ Department of Energy Office of Science and
the National Nuclear Security Administration, and the U.S. Department of
Energy, Office of Science, under Contract DE-AC02-06CH11357.
\end{acks}

\bibliographystyle{ACM-Reference-Format}
\bibliography{references}


\begin{thebibliography}{16}


\ifx \showCODEN    \undefined \def \showCODEN     #1{\unskip}     \fi
\ifx \showDOI      \undefined \def \showDOI       #1{#1}\fi
\ifx \showISBNx    \undefined \def \showISBNx     #1{\unskip}     \fi
\ifx \showISBNxiii \undefined \def \showISBNxiii  #1{\unskip}     \fi
\ifx \showISSN     \undefined \def \showISSN      #1{\unskip}     \fi
\ifx \showLCCN     \undefined \def \showLCCN      #1{\unskip}     \fi
\ifx \shownote     \undefined \def \shownote      #1{#1}          \fi
\ifx \showarticletitle \undefined \def \showarticletitle #1{#1}   \fi
\ifx \showURL      \undefined \def \showURL       {\relax}        \fi
\providecommand\bibfield[2]{#2}
\providecommand\bibinfo[2]{#2}
\providecommand\natexlab[1]{#1}
\providecommand\showeprint[2][]{arXiv:#2}

\bibitem[Ashraf et~al\mbox{.}(2018)]%
        {hetero-18}
\bibfield{author}{\bibinfo{person}{M~Usman Ashraf},
  \bibinfo{person}{Fathy~Alburaei Eassa}, \bibinfo{person}{Aiiad~Ahmad
  Albeshri}, {and} \bibinfo{person}{Abdullah Algarni}.}
  \bibinfo{year}{2018}\natexlab{}.
\newblock \showarticletitle{Performance and power efficient massive parallel
  computational model for {HPC} heterogeneous exascale systems}.
\newblock \bibinfo{journal}{\emph{IEEE Access}}  \bibinfo{volume}{6}
  (\bibinfo{year}{2018}), \bibinfo{pages}{23095--23107}.
\newblock


\bibitem[Balay et~al\mbox{.}(2023)]%
        {petsc}
\bibfield{author}{\bibinfo{person}{Satish Balay}, \bibinfo{person}{Shrirang
  Abhyankar}, \bibinfo{person}{Mark~F. Adams}, \bibinfo{person}{Steven Benson},
  \bibinfo{person}{Jed Brown}, \bibinfo{person}{Peter Brune},
  \bibinfo{person}{Kris Buschelman}, \bibinfo{person}{Emil Constantinescu},
  \bibinfo{person}{Lisandro Dalcin}, \bibinfo{person}{Alp Dener},
  \bibinfo{person}{Victor Eijkhout}, \bibinfo{person}{Jacob Faibussowitsch},
  \bibinfo{person}{William~D. Gropp}, \bibinfo{person}{V\'{a}clav Hapla},
  \bibinfo{person}{Tobin Isaac}, \bibinfo{person}{Pierre Jolivet},
  \bibinfo{person}{Dmitry Karpeev}, \bibinfo{person}{Dinesh Kaushik},
  \bibinfo{person}{Matthew~G. Knepley}, \bibinfo{person}{Fande Kong},
  \bibinfo{person}{Scott Kruger}, \bibinfo{person}{Dave~A. May},
  \bibinfo{person}{Lois~Curfman McInnes}, \bibinfo{person}{Richard~Tran Mills},
  \bibinfo{person}{Lawrence Mitchell}, \bibinfo{person}{Todd Munson},
  \bibinfo{person}{Jose~E. Roman}, \bibinfo{person}{Karl Rupp},
  \bibinfo{person}{Patrick Sanan}, \bibinfo{person}{Jason Sarich},
  \bibinfo{person}{Barry~F. Smith}, \bibinfo{person}{Stefano Zampini},
  \bibinfo{person}{Hong Zhang}, \bibinfo{person}{Hong Zhang}, {and}
  \bibinfo{person}{Junchao Zhang}.} \bibinfo{year}{2023}\natexlab{}.
\newblock \bibinfo{booktitle}{\emph{{PETSc/TAO} Users Manual}}.
\newblock \bibinfo{type}{{T}echnical {R}eport} ANL-21/39 - Revision 3.19.
  \bibinfo{institution}{Argonne National Laboratory}.
\newblock


\bibitem[Bernholdt et~al\mbox{.}(2020)]%
        {ecp-usage-20}
\bibfield{author}{\bibinfo{person}{David~E Bernholdt}, \bibinfo{person}{Swen
  Boehm}, \bibinfo{person}{George Bosilca}, \bibinfo{person}{Manjunath
  Gorentla~Venkata}, \bibinfo{person}{Ryan~E Grant}, \bibinfo{person}{Thomas
  Naughton}, \bibinfo{person}{Howard~P Pritchard}, \bibinfo{person}{Martin
  Schulz}, {and} \bibinfo{person}{Geoffroy~R Vallee}.}
  \bibinfo{year}{2020}\natexlab{}.
\newblock \showarticletitle{{A survey of MPI usage in the US Exascale Computing
  Project}}.
\newblock \bibinfo{journal}{\emph{Concurrency and Computation: Practice and
  Experience}} \bibinfo{volume}{32}, \bibinfo{number}{3}
  (\bibinfo{year}{2020}), \bibinfo{pages}{e4851}.
\newblock


\bibitem[Buntinas et~al\mbox{.}(2006)]%
        {nemesis-06}
\bibfield{author}{\bibinfo{person}{Darius Buntinas}, \bibinfo{person}{Guillaume
  Mercier}, {and} \bibinfo{person}{William Gropp}.}
  \bibinfo{year}{2006}\natexlab{}.
\newblock \showarticletitle{Design and evaluation of {Nemesis,} a scalable,
  low-latency, message-passing communication subsystem}. In
  \bibinfo{booktitle}{\emph{Sixth IEEE International Symposium on Cluster
  Computing and the Grid (CCGRID'06)}}, Vol.~\bibinfo{volume}{1}. IEEE,
  \bibinfo{pages}{10--pp}.
\newblock


\bibitem[Dinan et~al\mbox{.}(2014)]%
        {endpoints-14}
\bibfield{author}{\bibinfo{person}{James Dinan}, \bibinfo{person}{Ryan~E
  Grant}, \bibinfo{person}{Pavan Balaji}, \bibinfo{person}{David Goodell},
  \bibinfo{person}{Douglas Miller}, \bibinfo{person}{Marc Snir}, {and}
  \bibinfo{person}{Rajeev Thakur}.} \bibinfo{year}{2014}\natexlab{}.
\newblock \showarticletitle{Enabling communication concurrency through flexible
  {MPI} endpoints}.
\newblock \bibinfo{journal}{\emph{The International Journal of High Performance
  Computing Applications}} \bibinfo{volume}{28}, \bibinfo{number}{4}
  (\bibinfo{year}{2014}), \bibinfo{pages}{390--405}.
\newblock


\bibitem[Ferrer et~al\mbox{.}(2010)]%
        {OMP-offload-10}
\bibfield{author}{\bibinfo{person}{Roger Ferrer}, \bibinfo{person}{Vicen{\c{c}}
  Beltran}, \bibinfo{person}{Marc Gonzalez}, \bibinfo{person}{Xavier
  Martorell}, {and} \bibinfo{person}{Eduard Ayguad{\'e}}.}
  \bibinfo{year}{2010}\natexlab{}.
\newblock \showarticletitle{Analysis of task offloading for accelerators}. In
  \bibinfo{booktitle}{\emph{High Performance Embedded Architectures and
  Compilers: 5th International Conference, HiPEAC 2010, Pisa, Italy, January
  25-27, 2010. Proceedings 5}}. Springer, \bibinfo{pages}{322--336}.
\newblock


\bibitem[Hensgen et~al\mbox{.}(1988)]%
        {dissemination-88}
\bibfield{author}{\bibinfo{person}{Debra Hensgen}, \bibinfo{person}{Raphael
  Finkel}, {and} \bibinfo{person}{Udi Manber}.}
  \bibinfo{year}{1988}\natexlab{}.
\newblock \showarticletitle{Two algorithms for barrier synchronization}.
\newblock \bibinfo{journal}{\emph{International Journal of Parallel
  Programming}}  \bibinfo{volume}{17} (\bibinfo{year}{1988}),
  \bibinfo{pages}{1--17}.
\newblock


\bibitem[Hori et~al\mbox{.}(2018)]%
        {pip-18}
\bibfield{author}{\bibinfo{person}{Atsushi Hori}, \bibinfo{person}{Min Si},
  \bibinfo{person}{Balazs Gerofi}, \bibinfo{person}{Masamichi Takagi},
  \bibinfo{person}{Jai Dayal}, \bibinfo{person}{Pavan Balaji}, {and}
  \bibinfo{person}{Yutaka Ishikawa}.} \bibinfo{year}{2018}\natexlab{}.
\newblock \showarticletitle{Process-in-process: techniques for practical
  address-space sharing}. In \bibinfo{booktitle}{\emph{Proceedings of the 27th
  International Symposium on High-Performance Parallel and Distributed
  Computing}}. \bibinfo{pages}{131--143}.
\newblock


\bibitem[Kamal and Wagner(2010)]%
        {fg-mpi}
\bibfield{author}{\bibinfo{person}{Humaira Kamal} {and} \bibinfo{person}{Alan
  Wagner}.} \bibinfo{year}{2010}\natexlab{}.
\newblock \showarticletitle{{FG-MPI}: Fine-grain {MPI} for multicore and
  clusters}. In \bibinfo{booktitle}{\emph{2010 IEEE International Symposium on
  Parallel \& Distributed Processing, Workshops and Phd Forum (IPDPSW)}}.
  \bibinfo{pages}{1--8}.
\newblock
\urldef\tempurl%
\url{https://doi.org/10.1109/IPDPSW.2010.5470773}
\showDOI{\tempurl}


\bibitem[Krawezik(2003)]%
        {MPIvOMP-03}
\bibfield{author}{\bibinfo{person}{G{\'e}raud Krawezik}.}
  \bibinfo{year}{2003}\natexlab{}.
\newblock \showarticletitle{{Performance comparison of MPI and three OpenMP
  programming styles on shared memory multiprocessors}}. In
  \bibinfo{booktitle}{\emph{Proceedings of the fifteenth annual ACM symposium
  on Parallel algorithms and architectures}}. \bibinfo{pages}{118--127}.
\newblock


\bibitem[Lange et~al\mbox{.}(2013)]%
        {lange2013achieving}
\bibfield{author}{\bibinfo{person}{Michael Lange}, \bibinfo{person}{Gerard
  Gorman}, \bibinfo{person}{Michele Weiland}, \bibinfo{person}{Lawrence
  Mitchell}, {and} \bibinfo{person}{James Southern}.}
  \bibinfo{year}{2013}\natexlab{}.
\newblock \showarticletitle{Achieving efficient strong scaling with {PETSc
  using hybrid MPI/OpenMP} optimisation}. In
  \bibinfo{booktitle}{\emph{Supercomputing: 28th International Supercomputing
  Conference, ISC 2013, Leipzig, Germany, June 16-20, 2013. Proceedings 28}}.
  Springer, \bibinfo{pages}{97--108}.
\newblock


\bibitem[{Message Passing Interface Forum}(2021)]%
        {MPI4-threads}
\bibfield{author}{\bibinfo{person}{{Message Passing Interface Forum}}.}
  \bibinfo{year}{2021}\natexlab{}.
\newblock \bibinfo{title}{{MPI: A Message-Passing Interface Standard, Version
  4.0} - chap. 3.5 Semantics of Point-to-Point Communication}.
\newblock , \bibinfo{numpages}{54}~pages.
\newblock
\newblock
\shownote{\url{https://www.mpi-forum.org/docs/}}.


\bibitem[Ouyang et~al\mbox{.}(2020)]%
        {cabmpi-20}
\bibfield{author}{\bibinfo{person}{Kaiming Ouyang}, \bibinfo{person}{Min Si},
  \bibinfo{person}{Atsushi Hori}, \bibinfo{person}{Zizhong Chen}, {and}
  \bibinfo{person}{Pavan Balaji}.} \bibinfo{year}{2020}\natexlab{}.
\newblock \showarticletitle{{CAB-MPI: Exploring interprocess work-stealing
  towards balanced MPI communication}}. In \bibinfo{booktitle}{\emph{SC20:
  International Conference for High Performance Computing, Networking, Storage
  and Analysis}}. IEEE, \bibinfo{pages}{1--15}.
\newblock


\bibitem[P{\'e}rache et~al\mbox{.}(2009)]%
        {mpc-mpi}
\bibfield{author}{\bibinfo{person}{Marc P{\'e}rache}, \bibinfo{person}{Patrick
  Carribault}, {and} \bibinfo{person}{Herv{\'e} Jourdren}.}
  \bibinfo{year}{2009}\natexlab{}.
\newblock \showarticletitle{{MPC-MPI}: An {MPI} implementation reducing the
  overall memory consumption}. In \bibinfo{booktitle}{\emph{Recent Advances in
  Parallel Virtual Machine and Message Passing Interface: 16th European PVM/MPI
  Users’ Group Meeting, Espoo, Finland, September 7-10, 2009. Proceedings
  16}}. Springer, \bibinfo{pages}{94--103}.
\newblock


\bibitem[Zambre et~al\mbox{.}(2020)]%
        {Rohit-20}
\bibfield{author}{\bibinfo{person}{Rohit Zambre}, \bibinfo{person}{Aparna
  Chandramowliswharan}, {and} \bibinfo{person}{Pavan Balaji}.}
  \bibinfo{year}{2020}\natexlab{}.
\newblock \showarticletitle{How {I} learned to stop worrying about user-visible
  endpoints and love {MPI}}. In \bibinfo{booktitle}{\emph{Proceedings of the
  34th ACM International Conference on Supercomputing}}.
  \bibinfo{pages}{1--13}.
\newblock


\bibitem[Zhou et~al\mbox{.}(2022)]%
        {mpix-stream}
\bibfield{author}{\bibinfo{person}{Hui Zhou}, \bibinfo{person}{Ken Raffenetti},
  \bibinfo{person}{Yanfei Guo}, {and} \bibinfo{person}{Rajeev Thakur}.}
  \bibinfo{year}{2022}\natexlab{}.
\newblock \showarticletitle{{MPIX Stream}: An Explicit Solution to Hybrid
  {MPI+X} Programming}. In \bibinfo{booktitle}{\emph{Proceedings of the 29th
  European MPI Users' Group Meeting}}. \bibinfo{pages}{1--10}.
\newblock


\end{thebibliography}

\end{document}